\documentclass[prd,nofootinbib,onecolumn,12pt]{revtex4}
\usepackage{amsmath}
\usepackage{amssymb}
\usepackage{graphics}
\usepackage{epsfig}

\newcommand\nn{\nonumber}
\newcommand\ba{\begin{eqnarray}}
\newcommand\ea{\end{eqnarray}}
\newcommand{\br}[1]{\left( #1 \right)}
\newcommand{\brs}[1]{\left[ #1 \right]}

\newcommand{\brm}[1]{\left| #1 \right|}
\newcommand{\GeV}{~\mbox{GeV}}
\newcommand{\MeV}{~\mbox{MeV}}

\newcommand{\nb}{~\mbox{nb}}
\newcommand{\re}{\mbox{Re}}

\begin{document}

\title{Photoproduction of scalar and pseudoscalar mesons on a lepton
within the local Nambu-Jona-Lasinio model}

\author{E.~Barto\v{s}}
\email{fyziebar@savba.sk}
\affiliation{Institute of Physics, Slovak Academy of Sciences, Bratislava}

\author{Yu.~M.~Bystritskiy}
\email{bystr@theor.jinr.ru}
\affiliation{Joint Institute for Nuclear Research, Dubna, Russia}

\author{E.~A.~Kuraev}
\email{kuraev@theor.jinr.ru}
\affiliation{Joint Institute for Nuclear Research, Dubna, Russia}

\author{M.~Se{\v c}ansk\'y}
\email{fyzimsec@savba.sk}
\affiliation{Institute of Physics, Slovak Academy of Sciences, Bratislava}

\author{M.~K.~Volkov}
\email{volkov@theor.jinr.ru}
\affiliation{Joint Institute for Nuclear Research, Dubna, Russia}

\date{\today}

\begin{abstract}
Using the description of the subprocess $\gamma\gamma^*\to S(P)$ in terms of
local Nambu-Jona-Lasinio model we calculate the cross sections of photoproduction
of scalar and pseudoscalar mesons in high energy photon-lepton collisions
processes. The dependence on the transversal momentum and the total cross sections
in Weizsaecker-Williams approximation are presented.
\end{abstract}


\maketitle

\section{Introduction}

Photo-production of mesons on a lepton are the cross channel for the processes
of the radiative
production of a single meson (scalar or pseudoscalar) in electron-positron
annihilation \cite{Bystritskiy:2007wq}.
Production of the pseudo-scalar mesons in photon-lepton collisions can be
considered as a experiment of kind Primakoff one: photo-production of the neutral
pion in the Coulomb field of a heavy nuclei. It is rather difficult problem
to measure the Primakoff effect using as a target the field of proton. The reason
is a huge background effects connected with hadron structure of proton \cite{Pirmakoff:1951pj}.
When one consider the lepton as a target the background is absent.
We note that using as a target the electrons of a matter imply using very energetic
photons since the threshold photon energy is $E>M_\pi^2/(2m_e)\sim 20\GeV$.
Using the muon as a target  permits one to use the photons with energy of several
$\GeV$ when producing such a heavy mesons as $a_0(980)$, $f_0(980)$, $\sigma(600)$.

\section{The local Nambu-Jona-Lasinio model}

The local Nambu-Jona-Lasinio (NJL) model describes the
interaction of quarks with mesons by the Lagrangian
\cite{Volkov:1986zb,Volkov:2006vq}:
\ba
    {\cal L}_{int} &=&
    \bar q \left[ e Q \hat A +
    g_{\sigma_u} \lambda_u \sigma_u + g_{\sigma_s} \lambda_s \sigma_s + g_{\sigma_u} \lambda_3 a_0
    +\right. \nn\\
    &&\qquad+
        i \gamma_5 g_\pi \br{\lambda_{\pi^+} \pi^+ + \lambda_{\pi^-} \pi^- + \lambda_{3} \pi^0}
        +
        i \gamma_5 g_K \br{\lambda_{K^+} K^+ + \lambda_{K^-} K^-}
    +
    \nn\\
    &&\qquad+\left.
        i \gamma_5 \br{g_\pi \lambda_u \eta_u + g_{\eta_s} \lambda_s \eta_s}
    \right] q,
    \label{QuarkMesonLagrangian}
\ea
where $\bar q= \br{\bar u,\bar d,\bar s}$, and $u$, $d$, $s$ are the quark fields,
$Q=\mbox{diag}\br{2/3,-1/3,-1/3}$ is the quark electric charge matrix,
$e$ is the elementary electric charge ($e^2/4\pi=\alpha=1/137$),
$\lambda_u=\br{\sqrt{2} \lambda_0+\lambda_8}/\sqrt{3}$,
$\lambda_s=\br{-\lambda_0+\sqrt{2}\lambda_8}/\sqrt{3}$,
$\lambda_{\pi^\pm}=(\lambda_1\pm i\lambda_2)/\sqrt{2}$,
$\lambda_{K^\pm}=(\lambda_4\pm i\lambda_5)/\sqrt{2}$
where $\lambda_i$ are the well-known Gell-Mann matrices and
$\lambda_0 = \sqrt{2/3}~\mbox{diag}\br{1,1,1}$.
The coupling constants from the Lagrangian (\ref{QuarkMesonLagrangian}) are
defined in the following way \cite{Volkov:1986zb}:
\ba
g_{\sigma_u} &=& \left( 4 I^\Lambda\br{m_u, m_u}\right)^{-1/2} = 2.43, \nn\\
g_{\sigma_s} &=& \left( 4 I^\Lambda\br{m_s, m_s}\right)^{-1/2} = 2.99, \nn \\
g_{\pi} &=& \frac{m_u}{F_\pi} = 2.84, \nn\\
g_K &=& \frac{m_u+m_s}{2 F_K} = 3.01, \nn\\
g_{\eta_s} &=& \frac{m_s}{F_s} = 3.37, \nn
\ea
where $m_u=m_d=263\MeV$, $m_s=406\MeV$ are the constituent quark masses,
for $g_{\pi}$ and $g_K$ constants we used the Goldberger-Treiman relation,
$F_\pi=92.5\MeV$, $F_s = 1.3~F_\pi$ and $F_K = 1.2~F_\pi$, and $I^\Lambda\br{m, m}$ is the
logarithmically divergent integral which has the form:
\ba
    I(m,m) &=& \frac{N_c}{\br{2\pi}^4}
    \int d^4 k
    \frac{\theta\br{\Lambda^2-k^2}}
    {\br{k^2+m^2}^2} =
    \frac{N_c}{\br{4\pi}^2}
    \br{
        \ln\br{\frac{\Lambda^2}{m^2}+1}
        -
        \frac{\Lambda^2}{\Lambda^2 + m^2}
    }, \qquad N_c = 3. \nn
\ea
This integral is written in the Euclidean space.
The cut-off parameter $\Lambda = 1.27\GeV$
and constituent quark masses were taken
from \cite{Volkov:1986zb,Bystritskiy:2007wq}.

The scalar isoscalar mesons $f_0$, $\sigma$ are
the mixtures of pure $u$,$d$-quarks scalar state $\sigma_u$ and
pure $s$-quark scalar state $\sigma_s$:
\ba
f_0    &=& \sigma_u\sin\alpha + \sigma_s\cos\alpha, \nn \\
\sigma &=& \sigma_u\cos\alpha - \sigma_s\sin\alpha,
\ea
where mixing angle is $\alpha=11.3^o$ \cite{Volkov:1999qb,Volkov:1999qn,Volkov:2006vq}.
The pseudoscalar mesons $\eta$, $\eta'$ are
the mixtures of pure $u$,$d$-quarks scalar state $\eta_u$ and
pure $s$-quark scalar state $\eta_s$:
\ba
\eta  &=& -\eta_u\sin\theta + \eta_s\cos\theta, \nn \\
\eta' &=& \eta_u\cos\theta  + \eta_s\sin\theta,
\ea
where mixing angle is $\theta=51.3^o$ \cite{Volkov:1998ax,Volkov:1986zb}.

\section{General consideration of photoproduction processes}

We will consider the processes of photoproduction of
scalar and pseudoscalar mesons on lepton target
\ba
\gamma\br{k}+\mu\br{P} &\to& S\br{p_s}    +\mu\br{P'},
\qquad
S = a_0, f_0, \sigma, \nn\\
\gamma\br{k}+\mu\br{P} &\to& P\br{p_{p}}+\mu\br{P'},
\qquad
P = \pi_0, \eta, \eta', \nn
\ea
\ba
k^2 = 0, \quad P^2 = P^{'2} = m^2, \quad p_s^2 = M_S^2, \quad p_{p}^2 = M_{P}^2,
\quad s = 2 \br{kP} = 2 m E,
\ea
where $E$ is the energy of incident photon.
Within the local NJL model matrix elements of these processes have a form:
\ba
{\cal M}^{\gamma \mu \to S \mu} &=&
\frac{(4\pi\alpha)^{3/2}}{2\pi^2}
J_{\alpha}\br{P,P'} \frac{g^{\alpha\mu}}{q^2}
\epsilon^\nu(k)
\br{g_{\mu\nu}\br{k q}-k_\mu q_\nu}\rho^S, \\
{\cal M}^{\gamma \mu \to P \mu} &=&
\frac{(4\pi\alpha)^{3/2}}{2\pi^2}
J_{\alpha}\br{P,P'} \frac{g^{\alpha\mu}}{q^2}
\epsilon^\nu(k)
\br{\nu,k,\mu,q} \rho^{P},
\ea
where we use the notation $\br{a,b,c,d} = \varepsilon^{\alpha\beta\gamma\delta}
a_\alpha b_\beta c_\gamma d_\delta$,
$q=P-P'$ is the transferred momentum and
$J^\mu\br{P,P'}=\bar{u}\br{P'}\gamma^\mu u\br{P}$
is the electromagnetic current of lepton target.
The quantities $\rho^{S,P}$ encodes the contributions
of Feynman diagrams both with the quark loops and meson loops
and will be considered in Section~\ref{DefiniteChannels}.

We consider these processes in the peripheral kinematics
($s \gg q^2$), which gives
the main contribution to the total cross section.
For this aim we introduce the light-like auxiliary 4-vector
\ba
\tilde{P} = P-k\frac{m^2}{s},
\ea
\ba
\tilde{P}^2=0, \quad 2\tilde{P}P=m^2,
\quad 2\tilde{P}k=s, \nn
\ea
and use Sudakov decomposition of transferred momentum 4-vector $q$:
\ba
q=\alpha\tilde{P}+\beta k+q_\bot,
\qquad
q_\bot^2=-\vec{q}^2<0,
\ea
where $\vec{q}$ is the Euclidean  two-dimensional vector perpendicular to two
light-like vectors $k$, $\tilde{P}$ ($q_\bot k=q_\bot \tilde{P}=0$).
The square of the 4-vector of the transfer momentum $q$, using the on mass shell conditions
of the target lepton and the scalar (pseudoscalar) mesons can be written in form
\ba
q^2=-\frac{\vec{q}^2+m^2\alpha^2}{1-\alpha}\approx - \br{\vec{q}^2+q_{min}^2},
\qquad
q_{min}^2=\frac{m^2M_{S,P}^4}{s^2}.
\ea

The phase volume of final particles then reads as:
\ba
d\Gamma &=&
\frac{d^3p_{s,p}}{2E_{s,p}}\frac{d^3P'}{2E'}\frac{1}{(2\pi)^2}
\delta^4\br{k+P-p_{s,p}-P'}= \nn \\
&=&
\frac{1}{\br{2\pi}^2}\delta^4\br{P+k-p_{s,p}-P'}\delta^4\br{P-q-P'}
d^4p_{s,p}d^4P' d^4q
\times \nn\\
&\times&
\delta\br{\br{k+q}^2-M_{S,P}^2}
\delta\br{\br{P-q}^2-m^2}= \nn \\
&=&
\frac{1}{\br{2\pi}^2}\frac{d^2q}{2s}d\br{s\alpha}d\br{s\beta}
\delta\br{s\beta\br{1-\alpha}+\vec{q}^2+m^2\alpha}
\delta\br{q^2+s\alpha-M_{S,P}^2}.
\ea
After integration over the Sudakov parameters $\alpha$, $\beta$ we obtain
\ba
d\Gamma=\frac{d^2q}{8\pi^2 s}.
\ea
Next standard step is to use the Gribov representation for the Green function
of the virtual photon omitting terms which not enhanced by $s$:
\ba
g^{\alpha\mu} \approx \frac{2}{s}\tilde{P}^\mu k^\alpha.
\ea
The other components of the metric tensor contribution vanish in the limit
$s \gg M_s^2$, which is implied in peripheral kinematics.
Then the square of matrix element modulus is:
\ba
\sum \brm{{\cal M}^{\gamma \mu \to (S,P) \mu}}^2 =
\frac{32\alpha^3s^2\vec{q}^2}{\pi(q^2)^2}\brm{\rho^{S,P}}^2.
\ea
The differential cross section then have a form:
\ba
    \frac{d\sigma^{\gamma \mu \to (S,P) \mu}}{d\vec q^2} =
    \frac{\alpha^3 \vec q^2}
    {2\pi^2 \br{\vec{q}^2+q_{min}^2}^2} \brm{\rho^{S,P}}^2.
    \label{DiffCrossSection}
\ea

\section{Definite channels}
\label{DefiniteChannels}

Now we can calculate quantities $\rho^{S,P}$ for different channels of
photoproduction of different mesons.
The quantities $\rho^{S}$ have the contributions from the quark and the meson loops
and $\rho^{P}$ have only the contributions from quark loops.
Let us consider first process of $a_0$ meson production $\gamma+\mu\to a_0(980)+\mu$.
Charge-color factor associated with the $u$, $d$ quarks is $3((4/9)+1/9)=5/3$. The result is
\ba
\rho^{a_0}=\frac{5g_{\sigma_u}}{3m_u}I^{a_0}_u+\frac{g_{a_0K^+K^-}}{M_K^2}I^{a_0}_K,
\ea
where integrals corresponding to quark and meson loops are \cite{Volkov:2009mz}
\ba
I^S_q &=& \re\br{\int\limits_0^1dx\int\limits_0^{1-x}
\frac{dz\br{-1+4xz}}{1-zx\frac{M_S^2}{m_q^2}+xy\frac{\vec{q}^2}{m_q^2}}},
\qquad y=1-x-z, \nn \\
I^S_M &=& \int\limits_0^1dx\int\limits_0^{1-x}
\frac{dz\br{-xz}}{1-zx\frac{M_S^2}{M^2}+xy\frac{\vec{q}^2}{M^2}},
\ea
where $m_q$ and $M$ are the masses of quarks and mesons circulating in the loop.

For process $\gamma+\mu\to f_0(980)+\mu$ we obtain:
\ba
\rho^{f_0} &=&
\brs{
    \frac{5g_{\sigma_u}}{3m_u}I^{f_0}_u-
    4\frac{m_u}{M_\pi^2}\frac{g_\pi^2}{g_{\sigma_u}}I_\pi^{f_0}+
    \frac{g_{a_0K^+K^-}}{M_K^2}I^{f_0}_K
}\sin\alpha+\nn \\
&+&
\brs{
    \frac{1}{3}\frac{g_{\sigma_s}}{m_s} I^{f_0}_s +
    \frac{g_{\sigma_s K^+K^-}}{M_K^2}I^{f_0}_K
} \cos\alpha
\ea
and finally for $\gamma+\mu\to \sigma(600)+\mu$ we obtain:
\ba
\rho^{\sigma(600)} &=&
\brs{
    \frac{5g_{\sigma_u}}{3m_u}I^{f_0}_u-
    4\frac{m_u}{M_\pi^2}\frac{g_\pi^2}{g_{\sigma_u}}I_\pi^{f_0}+
    \frac{g_{a_0K^+K^-}}{M_K^2}I^{a_0}_K
}\cos\alpha-\nn \\
&-&
\brs{
    \frac{1}{3}\frac{g_{\sigma_s}}{m_s}I^{f_0}_s+
    \frac{g_{\sigma_sK^+K^-}}{M_K^2}I^{a_0}_K
}\sin\alpha.
\ea
The defined channels of the photoproduction of
pseudoscalars $P = \pi^0, \eta, \eta'$ have a form:
\ba
    \rho^{\pi^0}
    &=&
    \frac{5}{3} \frac{1}{F_\pi} J^{\pi^0}_{u}, \\
    \rho^{\eta}
    &=&
        -\sin\theta \frac{5}{3} \frac{1}{F_\pi} J^{\eta}_{u}
        +\cos\theta \frac{1}{3} \frac{1}{F_s}   J^{\eta}_{s}, \\
    \rho^{\eta'}
    &=&
         \cos\theta \frac{5}{3} \frac{1}{F_\pi} J^{\eta'}_{u}
        +\sin\theta \frac{1}{3} \frac{1}{F_s}   J^{\eta'}_{s},
\ea
where loop integrals $J^P_q$ have the form:
\ba
J^P_q &=& \re\br{\int\limits_0^1dx\int\limits_0^{1-x}
\frac{dz}{1-zx\frac{M_P^2}{m_q^2}+xy\frac{\vec{q}^2}{m_q^2}}},
\qquad y=1-x-z.
\ea

\begin{figure}
\includegraphics[width=0.8\textwidth]{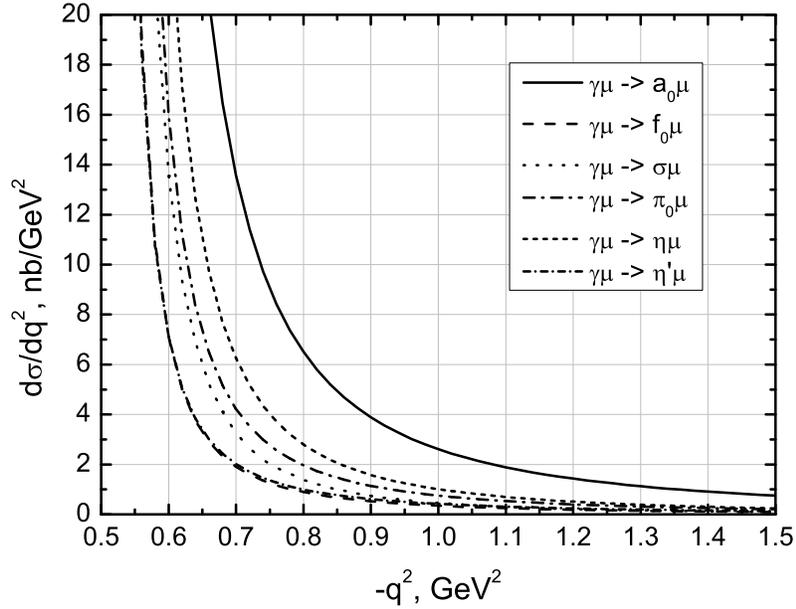}
\caption{The dependence of cross sections $\frac{d\sigma^{\gamma \mu \to (S,P) \mu}}{d\vec q^2}$
(\ref{DiffCrossSection})
for different channels of transfer momentum square $\vec q^2$.
\label{Plots}}
\end{figure}

The dependence of cross sections for different channels of
transfer momentum square $\vec q^2$ is shown in Fig.~\ref{Plots}.

The total cross section in the Weizsaecker-Williams approximation
have the form:
\ba
    \sigma^{\gamma \mu \to (S,P) \mu}
    =
    \frac{\alpha^3 L}{2\pi^2} \brm{\rho^{S,P}\br{0}}^2,
    \qquad
    L = 2\ln \br{\frac{q_{max}^2}{q_{min}^2}} =
    2\ln \br{\frac{s}{m M}} \approx 10,
    \qquad
    s \approx 5\GeV^2,
\ea
where $L$ is the big logarithm which enhance the
cross section in peripheric kinematic.
The matrix element square $\rho^{S,P}\br{0}$
in this approximation for different channels have a form:
\ba
    \begin{array}{lcl}
        \rho^{\pi^0}\br{0} = -9.22\GeV^{-1},
        &\qquad&
        \rho^{a^0}\br{0} = -10.6\GeV^{-1}, \\
        \rho^{\eta}\br{0} = 14.4\GeV^{-1},
        &\qquad&
        \rho^{\sigma}\br{0} = -12.5\GeV^{-1}, \\
        \rho^{\eta'}\br{0} = -3.52\GeV^{-1},
        &\qquad&
        \rho^{f^0}\br{0} = -5.14\GeV^{-1}. \\
    \end{array}
\ea
%

\section{Results and discussion}

The total cross sections of processes $\gamma \mu\to (S,P)\mu$
in the Weizsaecker-Williams approximation for $\sqrt{s} > 3\GeV$
are:
\ba
    \begin{array}{lclcl}
        \sigma^{\gamma \mu\to \pi^0\mu} = 5.6\nb,
        &\qquad&
        \sigma^{\gamma \mu\to \eta\mu} = 13.8\nb,
        &\qquad&
        \sigma^{\gamma \mu\to \eta'\mu} = 0.8\nb,
        \\
        \sigma^{\gamma \mu\to a_0\mu} = 7.6\nb,
        &\qquad&
        \sigma^{\gamma \mu\to \sigma\mu} = 10.4\nb,
        &\qquad&
        \sigma^{\gamma \mu\to f_0\mu} = 1.7\nb.
    \end{array}
\ea

The similar estimations will be valid for proton or
nuclei instead of muon. The question about the background
events is important in this case.
The background will be mainly provided by other mechanisms
of meson production when the initial photon interacts
directly with the target.
Kinematics of the processes considered above is quite different
-
really the energy of produced mesons is large
(of order of the initial photon) and the cross sections
practically do not depend on $\sqrt{s}$. These features of this
events can be used to suppress the background.

We note that our results do not agree with the asymptotic behavior of
transition formfactor obtained in \cite{Dorokhov:2002tq}.
Really in \cite{Dorokhov:2002tq} the modification of loop integrals
was done, i.e. the cut was introduced to avoid large logarithm
$\ln\br{q^2/M^2}$ type contribution. In our case the
characteristic momenta of loop integral is of order
$\brm{k^2} \sim \brm{q^2}$ and we do not put any cut.
For $\vec q^2 \to 0$ our results are in agreement with current
algebra ones.


\end{document}